\documentclass[12pt]{iopart}
\expandafter\let\csname equation*\endcsname\relax
\expandafter\let\csname endequation*\endcsname\relax
\usepackage{amsmath}
\usepackage{cite}
\usepackage{graphicx}
\usepackage{bm}% bold math
\usepackage{hyperref}% add hypertext capabilities超链接
\hypersetup{colorlinks=true, citecolor=blue, urlcolor=blue, linkcolor=blue}
\usepackage{subfigure}
\begin{document}

\title[]{Vector rogue waves in spin-1 Bose-Einstein condensates with spin-orbit coupling}

\author{Jun-Tao~He\textsuperscript{1}, Hui-Jun~Li\textsuperscript{1}, Ji~Lin\textsuperscript{1,*}, and Boris A. Malomed\textsuperscript{2,3}}
\address{\textsuperscript{1} Department of Physics, Zhejiang Normal University, Jinhua 321004, China
\\ \textsuperscript{2} Department of Physical Electronics, School of Electrical Engineering, 
Faculty of Engineering, Tel Aviv University, Tel Aviv 69978, Israel
\\ \textsuperscript{3} Instituto de Alta Investigaci\'{o}n, Universidad de Tarapac\'{a}, Casilla 7D, Arica, Chile
\\ \textsuperscript{*} Authors to whom any correspondence should be addressed.}
\ead{linji@zjnu.edu.cn}
\vspace{10pt}

\begin{abstract}
We analytically and numerically study three-component rogue waves (RWs) in spin-1 Bose-Einstein condensates with Raman-induced spin-orbit coupling (SOC). Using the multiscale perturbative method, we obtain approximate analytical solutions for RWs with positive and negative effective masses, determined by the effective dispersion of the system. The solutions include RWs with smooth and striped shapes, as well as higher-order RWs. The analytical solutions demonstrate that the RWs in the three components of the system exhibit different velocities and their maximum peaks appear at the same spatiotemporal position, which is caused by SOC and interactions. The accuracy of the approximate analytical solutions is corroborated by comparison with direct numerical simulations of the underlying system. Additionally, we systematically explore existence domains for the RWs determined by the baseband modulational instability (BMI). Numerical simulations corroborate that, under the action of BMI, plane waves with random initial perturbations excite RWs, as predicted by the approximate analytical solutions.
\end{abstract}

\vspace{2pc}
\noindent{\it Keywords}: spin-orbit coupling, Bose-Einstein condensates, rogue waves, baseband modulational instability

%Uncomment for Submitted to journal title message
\submitto{\NJP}
%\maketitle

\section{Introduction}

Rogue waves (RWs), renowned for their extreme destructiveness and unpredictability in the ocean, have been a puzzling phenomenon for many years. The emergence of an analytical rational solution called Peregrine soliton was a starting point for the theoretical studies of RWs \cite{PeregrineDH1983Water}. Subsequently, mechanisms that underlie and control RWs have become important topics. RWs have been found theoretically and experimentally in many nonlinear media, including oceans \cite{DystheK2008Oceanic} and other realizations of hydrodynamics \cite{Shats2010Capillary,Chabchouba2011Rogue}, acoustics \cite{TsaiYY2016Generation,Joseph2021Modulational}, plasmas \cite{MoslemWM2011Surface,DingCC2020Lax,RaoJG2021Completely}, optics \cite{SolliDR2007Optical}, Bose-Einstein condensates (BECs) \cite{BludovVYu2009Matter,QinZ2012Matter,RomeroRosA2022Theoretical}, and even financial markets \cite{YanZY2010Financial,YanZY2011Vector}.

Recently, the experimental realization of RWs in repulsive two-component BECs has attracted much interest \cite{RomeroRosA2024Experimental}, which suggests a possibility to look for RWs in binary BEC with the spin-orbit coupling (SOC) between the components \cite{LinYJ2011Spin,LanZ2014Raman,LuoX2016Tunable}, which helps to construct a variety of stable bound states \cite{WangC2010Spin,AdhikariSK2021Multiring,ZhangY2023Quantum,HeJ2023Stationary}. RWs in this system have been recently studied in works \cite{HeJT2022Multi,ChenYX2023Vector}. However, effects of some essential factors, such as the Zeeman splitting and Raman coupling, on RWs have not been addressed yet.

Analytical considerations of RWs in various systems rely upon a direct approach \cite{TajiriM1998Breather,AkhmedievN2009Rogue,ZhaoLC2013Rogue,HeJS2013Generating,BaronioF2014Vector,ChenS2015Optical,LiuC2022Non} or simplification provided by similarity transformations \cite{YanZ2010Three,WenL2011Matter,LoombaS2014Controlling,ManikandanK2014Manipulating,LiuC2014Optical,HeJS2014Rogue, ChabchoubA2014Time,ManikandanK2016Manipulating,KengneE2023Baseband,KengneE2023Manipulating}. As is well known, the excitation of RWs is closely related to the baseband modulation instability (BMI) of plane waves in the systems \cite{BludovVYu2010Vector,OnoratoM2013Rogue,BaronioF2015Baseband,LingL2017Generation,GelashAA2018Formation, PanC2021Omnipresent,ChenS2022Modulation,LiuL2023Formation}, including non-integrable ones \cite{ToengerS2015Emergent,GaoP2020High,TabiCB2021Generation,TanY2022Super}. In the latter case, the absence of exact solutions makes it necessary to use the multiscale perturbation method, properly combined with numerical simulations \cite{AchilleosV2013Matter,ZhaoLC2020Magnetic,LiXX2021Solitary,MithunT2024Stationary}.

In this work, we derive two types of approximate three-component (vector) RWs solutions in the spin-1 BEC system with the Raman-induced SOC, using the multiscale method. These are RWs with smooth and striped shapes. The corresponding higher-order RWs are obtained too. The SOC and Raman terms cause the three components of the vector RWs to exhibit different velocities and break the axial symmetry of the higher-order vector RWs. Numerical simulations reproduce these approximate vector RWs, indicating that the multiscale perturbation method is also suitable for producing nonstationary solutions. In addition, we find exact existence domains of these RWs, based on the BMI of plane waves, which is close to the prediction of the multiscale perturbation method. We also verify the RW existence domain numerically. In the BMI region, the plane waves with Gaussian perturbations excite the RW in the course of the numerical evolution, while RWs cannot be excited from modulationally stable plane waves.

This paper is organized as follows. The model of the spin-1 BECs with the Raman-induced SOC is introduced in Sec.~\ref{sec2}, where it is simplified into a single nonlinear Schr\"{o}dinger (NLS) equation. In Sec.~\ref{sec3}, we obtain and analyze analytical RW solutions in the system. In Sec.~\ref{sec4}, the existence domains of RWs are obtained from the consideration of BMI. The paper is concluded in Sec.~\ref{sec5}.

\section{The model and simplification}

\label{sec2} We consider a spin-1 BEC with the Raman-induced SOC, which is subject to tight confinement in the transverse directions with a large trapping frequency $\omega_{\perp}$. Accordingly, effective dynamics is reduced to the single longitudinal coordinate $x$. The corresponding single-particle Hamiltonian \cite{LanZ2014Raman} is
\begin{eqnarray}
H_{0}=\frac{\left( p_{x}+2\hbar k_{r}\Sigma_{z}\right)^{2}}{2m}+\frac{\hbar \Omega_{R}}{\sqrt{2}}\Sigma_{x}-\hbar \delta_{R}\Sigma_{z}+\hbar \varepsilon_{R}\Sigma_{z}^{2},
\end{eqnarray}
where $m$ is the atomic mass and $p_{x}=-\mathrm{i}\hbar \partial_{x}$ is the momentum operator along the direction of the Raman laser beams, $2\hbar k_{r}$ being the photon recoil momentum that determines the SOC strength. $\Omega_{R}$ and $\delta_{R}$ are the resonant Raman frequency and detuning, while $\hbar \varepsilon_{R}$ represents the quadratic Zeeman effect. Here, spin-1 matrices $\Sigma_{x}$ and $\Sigma_{z}$ are given in the commonly known irreducible representation by
\begin{eqnarray}
\Sigma_{x}=\frac{1}{\sqrt{2}}\left(
\begin{array}{lll}
0 & 1 & 0 \\
1 & 0 & 1 \\
0 & 1 & 0
\end{array}
\right) ,~~\Sigma_{z}=\left(
\begin{array}{ccc}
1 & 0 & 0 \\
0 & 0 & 0 \\
0 & 0 & -1
\end{array}
\right).
\end{eqnarray}
In the mean-field approximation, the spin-1 BEC with SOC is governed by the quasi-one-dimensional three-component Gross-Pitaevskii equations \cite{KawaguchiY2012Spinor}. Measuring the energy, length, and time in the units of $\hbar \omega_{\perp }$, $\sqrt{\hbar /(m\omega_{\perp })}$, and $\omega_{\perp}^{-1}$, respectively, the so rescaled equations become
\begin{eqnarray}
\begin{aligned}
\mathrm{i}\frac{\partial \psi_{\pm 1}}{\partial {t}}=&\left( -\frac{1}{2}\frac{\partial ^{2}}{\partial {x}^{2}}+V(x)\mp \mathrm{i}\gamma \frac{\partial }{\partial {x}}\mp \delta \right) \psi_{\pm 1}+\Omega \psi_{0} \\
&+\left( c_{0}\rho +c_{2}\left( \rho_{\pm 1}+\rho_{0}-\rho_{\mp 1}\right)\right) \psi_{\pm 1}+c_{2}\psi_{0}^{2}\psi_{\mp 1}^{\ast }, \\
\mathrm{i}\frac{\partial \psi_{0}}{\partial {t}}=&\left( -\frac{1}{2}\frac{\partial^{2}}{\partial {x}^{2}}+V(x)-\varepsilon \right) \psi_{0}+\Omega \left(
\psi_{+1}+\psi_{-1}\right) \\
&+\left( c_{0}\rho +c_{2}\left( \rho_{+1}+\rho_{-1}\right) \right) \psi_{0}+2c_{2}\psi_{0}^{\ast }\psi_{+1}\psi_{-1},  \label{eq3}
\end{aligned}
\end{eqnarray}
where the dimensionless parameters are $\gamma=2k_{r}\sqrt{\hbar /(m\omega_{\perp})}$, $\Omega=\Omega_{R}/(2\omega_{\perp})$, $\varepsilon=\varepsilon_{R}/\omega_{\perp }+2\hbar k_{r}^{2}/(m\omega_{\perp })$, and $\delta=\delta_{R}/\omega_{\perp }$. In addition, $\psi_{j}$ (with $j=\pm1,0$) are three components of the wave functions of the spinor BEC, and the longitudinal trapping potential, $V(x)$, is dropped below, as we aim to consider RW states in free space. Further, $\rho_{j}=|\psi_{j}|^{2}$ are the component densities, while $\rho=\sum_{j}|\psi_{j}|^{2}$ is the total particle density. Constants of the mean-field interaction $c_{0}$ and spin-exchange interaction $c_{2}$ are related to two-body \textit{s}-wave scattering lengths $a_{0}$ and $a_{2}$ for the total spin $0$ and $2$, respectively: $c_{0}=2\left( a_{0}+2a_{2}\right) /(3\sqrt{\hbar /(m\omega_{\perp})})$, $c_{2}=2\left( a_{2}-a_{0}\right) /(3\sqrt{\hbar /(m\omega_{\perp })})$. These constants, which can be adjusted experimentally by means of the Feshbach-resonance technique \cite{PapoularDJ2010Microwave}, have a strong impact on the types of nonlinear waves existing in the system.

Next, we attempt to simplify the non-integrable system \eqref{eq3} into an integrable NLS equation by means of the multiscale perturbation method. To this end, we adopt the solution as
\begin{eqnarray}
\boldsymbol{\psi}=\sum_{n=1}^{\infty}\epsilon^{n}\boldsymbol{u}_{n}\mathrm{e}^{\mathrm{i}(kx-\mu t)}\equiv \sum_{n=1}^{\infty}\epsilon^{n}\boldsymbol{\xi}_{n}\varphi_{n}(X,T)\mathrm{e}^{\mathrm{i}(kx-\mu t)}, \label{eq4}
\end{eqnarray}
where $\boldsymbol{\psi}=(\psi_{+1},\psi_{0},\psi_{-1})^{T}$, $\boldsymbol{\xi }_{n}=(U_{n},V_{n},W_{n})^{T}$ are sets of real coefficients, and $\varphi_{n}(X,T)$ are functions of the slow variables $T=\epsilon^{2}t$ and $X=\epsilon (x-vt)$ ($v$ is the group velocity of the carrier wave), $\epsilon \ll 1$ is a small parameter and $k$ is the momentum. Further, $\mu=\omega+\epsilon^{2}\omega_{2}$ is the chemical potential, $\omega$ is the single-particle energy, and $\epsilon^{2}\omega_{2}$ is a small correction to it. 

Substituting the ansatz~\eqref{eq4} in equation~\eqref{eq3}, we derive the following equations at orders $O(\epsilon )$, $O(\epsilon ^{2})$, and $O(\epsilon^{3})$, respectively:
\begin{eqnarray}
\boldsymbol{M}\boldsymbol{u}_{1} &=&0,  \label{eq5} \\
\boldsymbol{M}\boldsymbol{u}_{2} &=&\mathrm{i}\boldsymbol{Q}\partial_{X}\boldsymbol{u}_{1},  \label{eq6} \\
\boldsymbol{M}\boldsymbol{u}_{3} &=&\mathrm{i}\boldsymbol{Q}\partial_{X}\boldsymbol{u}_{2}-\left( \mathrm{i}\partial_{T}+\frac{1}{2}\partial_{X}^{2}-\boldsymbol{G}+\omega_{2}\right) \boldsymbol{u}_{1},  \label{eq7}
\end{eqnarray}
where matrices $\boldsymbol{M}$, $\boldsymbol{Q}$, and $\boldsymbol{G}$ are
\begin{eqnarray}
\begin{aligned}
\boldsymbol{M}&=\left(\begin{array}{ccc} M_{1} & \Omega & 0 \\ \Omega & M_{2} & \Omega \\ 0 & \Omega & M_{3}\end{array}\right), 
~\boldsymbol{Q}=\left(\begin{array}{ccc} Q_{1} & 0 & 0 \\ 0 & Q_{2} & 0 \\ 0 & 0 & Q_{3}\end{array}\right), \\ \boldsymbol{G}&=\left(\begin{array}{ccc} G_{1}& c_2V_1W_1 & 0 \\ c_2V_1W_1 & G_{2} & c_2U_1V_1 \\ 0 & c_2U_1V_1 & G_{3}\end{array}\right)|\varphi_1|^2.
\end{aligned}
\end{eqnarray}
with diagonal elements
\begin{eqnarray}
\begin{aligned}
M_{1}&=\frac{1}{2}k^2+\gamma k-\delta-\omega,\\ M_{2}&=\frac{1}{2}k^2-\varepsilon-\omega,~ M_{3}=\frac{1}{2}k^2-\gamma k+\delta-\omega,\\ Q_{1}&=k-v+\gamma,~Q_{2}=k-v,~Q_{3}=k-v-\gamma,\\ G_{1}&=c_0\left(U_1^2+V_1^2+W_1^2\right)+c_2\left(U_1^2+V_1^2-W_1^2\right), \\ G_{2}&=c_0\left(U_1^2+V_1^2+W_1^2\right)+c_2\left(U_1^2+W_1^2\right),\\
G_{3}&=c_0\left(U_1^2+V_1^2+W_1^2\right)+c_2\left(W_1^2+V_1^2-U_1^2\right).
\end{aligned}  \label{eq9}
\end{eqnarray}
At order $O(\epsilon)$, we obtain the single-particle energy spectrum from the solvability condition $\mathrm{det}(\boldsymbol{M})=0$. There are three different energy branches,
\begin{eqnarray}
\begin{aligned}
\omega_{+1}&=\frac{k^2}{2}-\frac{\varepsilon}{3}+\sqrt[3]{\mathrm{i}\sqrt{\Delta}-\Gamma}+\sqrt[3]{-\mathrm{i}\sqrt{\Delta}-\Gamma},\\
\omega_{0}&=\frac{k^2}{2}-\frac{\varepsilon}{3}+\alpha^*\sqrt[3]{\mathrm{i}\sqrt{\Delta}-\Gamma}+\alpha\sqrt[3]{-\mathrm{i}\sqrt{\Delta}-\Gamma},\\
\omega_{-1}&=\frac{k^2}{2}-\frac{\varepsilon}{3}+\alpha\sqrt[3]{\mathrm{i}\sqrt{\Delta}-\Gamma}+\alpha^*\sqrt[3]{-\mathrm{i}\sqrt{\Delta}-\Gamma}, \end{aligned}
\label{eq10}
\end{eqnarray}
where $\Delta=K^{2}\left(K^{2}-\varepsilon^{2}\right)^{2}+\Omega^{4}\varepsilon^{2}+6\Omega^{2}K^{4}+12\Omega^{4}K^{2}+10\varepsilon^{2}\Omega^{2}K^{2}+8\Omega^{6}$, $\Gamma=\varepsilon \left(\varepsilon^{2}+9\Omega^{2}-9K^{2}\right) /27$, $K=k\gamma-\delta$, and $\alpha=(-1+\sqrt{3}\mathrm{i})/2$. Note that the last two terms in each equation in system~\eqref{eq10} are complex conjugates of each other, thus ensuring that the energy spectrum remains real.

\begin{figure}[tbph]
	\centering
	\includegraphics[width=0.6\linewidth]{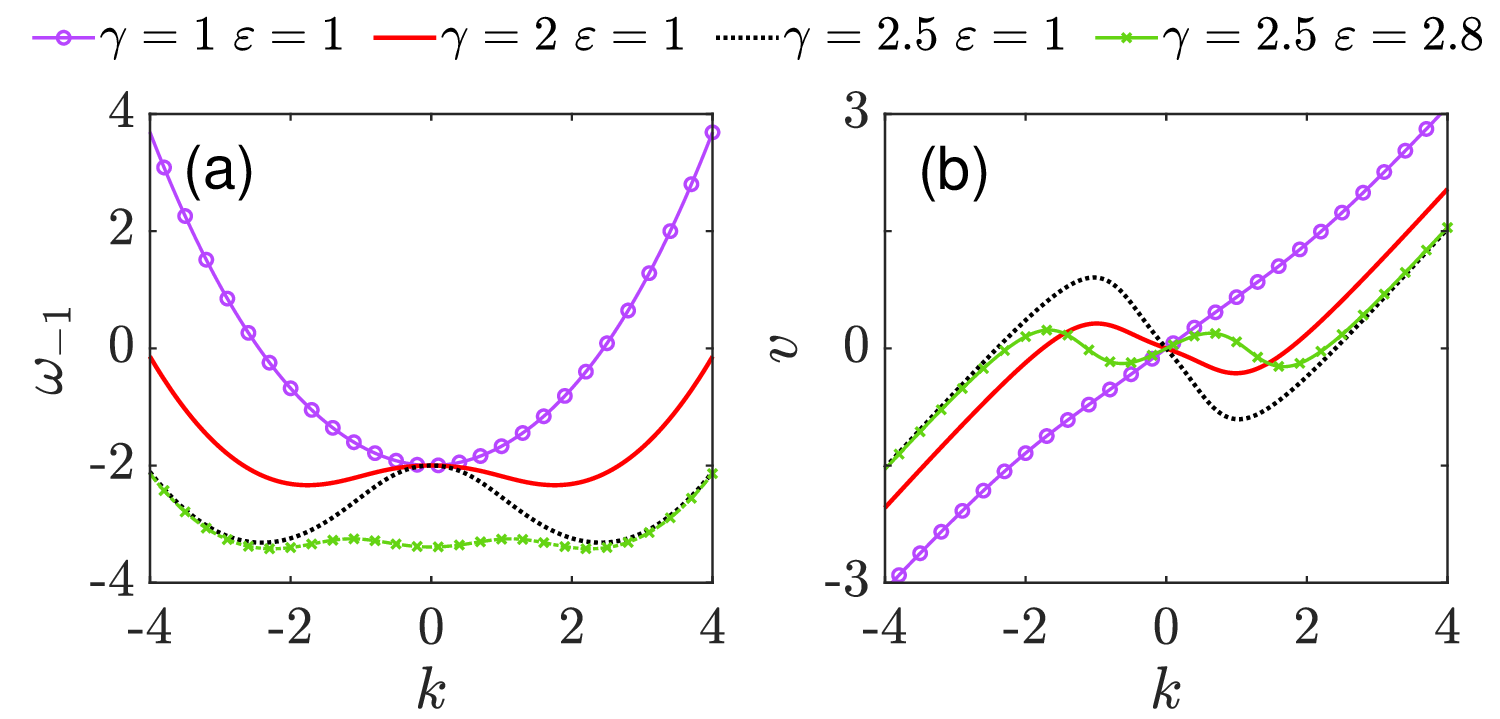}
	\caption{(a) The lowest energy branch $\protect\omega_{-1}$ and (b) the group velocity $v$ vs. momentum $k$ for $\Omega=1$, $\protect\delta=0$, and different values of $\protect\gamma$ and $\protect\varepsilon$.} \label{fig:lowest_energy_group_v}
\end{figure}

We focus on the lowest energy branch $\omega_{-1}$ in the momentum space $k$, as shown in figure~\ref{fig:lowest_energy_group_v}(a). By adjusting parameters, we can make this branch to have up to three local minima, corresponding $k=k_{\mathrm{min}}$, which determine the types of plane waves in the system.
Furthermore, we find that $\boldsymbol{\xi }_{1}$ is an eigenvector of $\boldsymbol{M}$ at the zero eigenvalue. Without loss of generality, we set $V_{1}=1$, the corresponding eigenvector being
\begin{eqnarray}
\boldsymbol{\xi }_{1}=(U_{1},V_{1},W_{1})^{T}=\left( -\frac{\Omega}{M_{1}},1,-\frac{\Omega}{M_{3}}\right)^{T}. \label{eq11}
\end{eqnarray}

The Hermitian matrix $\boldsymbol{M}$ has $\boldsymbol{\xi }_{1}^{T}\boldsymbol{M}=0$, and we thus obtain the compatibility condition $\boldsymbol{\xi }_{1}^{T}\boldsymbol{Q}\boldsymbol{\xi }_{1}=0$ at order $O(\epsilon^{2})$. The group velocity is given by the usual expression,
\begin{eqnarray}
v=\frac{\partial \omega }{\partial k}=k+\frac{\gamma \left(U_{1}^{2}-W_{1}^{2}\right) }{1+U_{1}^{2}+W_{1}^{2}},  \label{eq12}
\end{eqnarray}
as shown in figure~\ref{fig:lowest_energy_group_v}(b). Next, we obtain the following solution for $\boldsymbol{u}_{2}$ based on equations~\eqref{eq11} and \eqref{eq12}:
\begin{eqnarray}
\boldsymbol{u}_{2}=\boldsymbol{\xi }_{2}\varphi_{2}=-\frac{\mathrm{i}}{\Omega }\left(
\begin{array}{c}
\left( k-v+\gamma \right) U_{1}^{2} \\0 \\
\left( k-v-\gamma \right) W_{1}^{2}
\end{array}\right) \partial_{X}\varphi_{1}.  \label{eq13}
\end{eqnarray}

Finally, to address order $O(\epsilon^{3})$, we left multiply both sides of equation~\eqref{eq7} by $\boldsymbol{\xi}_{1}$. Combined with equations~\eqref{eq11} and \eqref{eq13}, another compatibility condition is obtained, which is the NLS equation for $\varphi_{1}$:
\begin{eqnarray}
i\frac{\partial \varphi_1}{\partial T}+\frac{1}{2}P\frac{\partial^{2}\varphi_{1}}{\partial X^{2}}+S\left\vert \varphi_{1}\right\vert^{2}\varphi_{1}+\omega_{2}\varphi_{1}=0,  \label{eq14}
\end{eqnarray}
where the coefficients of the effective dispersion and nonlinear interaction are, respectively,
\begin{eqnarray}
P=1+\frac{2\left( k-v+\gamma \right) ^{2}U_{1}^{3}+2\left( k-v-\gamma \right) ^{2}W_{1}^{3}}{\Omega \left( 1+U_{1}^{2}+W_{1}^{2}\right) }, \label{eq15}
\end{eqnarray}
\begin{eqnarray}
S=\frac{\left( 1-2U_{1}W_{1}\right)^{2}}{1+U_{1}^{2}+W_{1}^{2}}c_{2}-\left( 1+U_{1}^{2}+W_{1}^{2}\right) \left( c_{0}+c_{2}\right) .  \label{eq16}
\end{eqnarray}
The effective dispersion $P$ also represents the derivative of group velocity $v$ with respect to $k$.

According to equations~\eqref{eq4}, \eqref{eq11}, and \eqref{eq13}, we obtain the second-order approximate solution for the original equation~\eqref{eq3} in the form of
\begin{eqnarray}
\begin{aligned}
\boldsymbol{\psi}=&\epsilon \mathrm{e}^{\mathrm{i}kx-\mathrm{i}(\omega+\epsilon^2\omega_2)t}\left(\begin{array}{c}-\frac{\Omega}{M_1}-\frac{\mathrm{i}(k-v+\gamma)\Omega}{M_1^2}\partial_x \\ 1 \\
-\frac{\Omega}{M_3}-\frac{\mathrm{i}(k-v-\gamma)\Omega}{M_3^2}\partial_x
\end{array}\right)\varphi_1,
\end{aligned}  \label{eq17}
\end{eqnarray}
where $\varphi_{1}$ is any exact solution of the NLS equation (\ref{eq14}). Finally, we need to substitute $X=\epsilon (x-vt)$ and $T=\epsilon^{2}t$ in $\varphi_{1}\left(X,T\right) $ to obtain the approximate analytical solution of the original system.

\section{Vector rogue waves}

\label{sec3} Using equation~\eqref{eq17}, one can obtain multiple types of approximate analytical solutions for the spin-1 BEC system with SOC. We focus on RWs corresponding to the lowest energy branch $\omega =\omega_{-1}$, see equation~\eqref{eq10}.

\begin{figure}[htbp]
	\centering
	\includegraphics[width=0.7\linewidth]{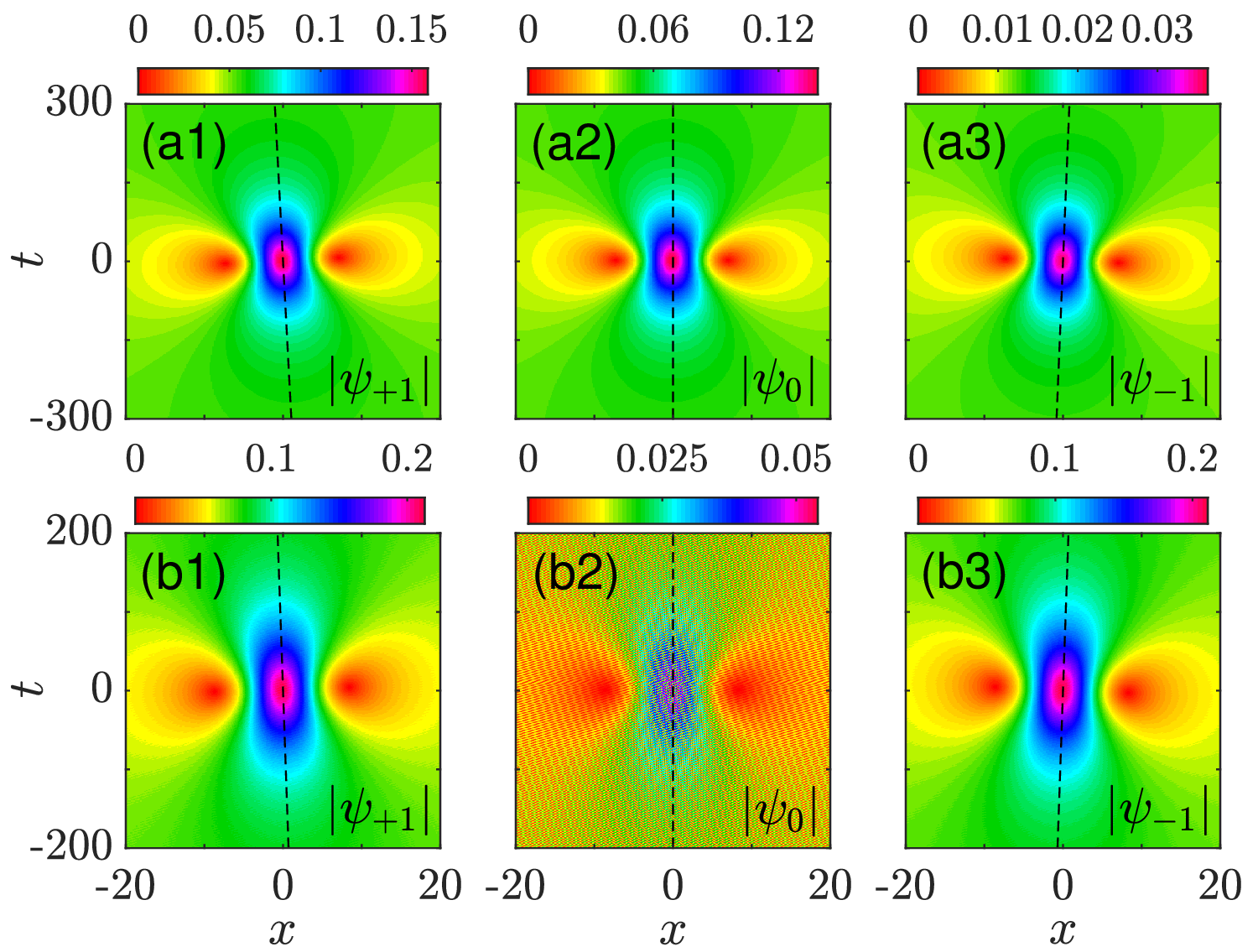}
	\caption{The density profiles of the positive-mass first-order smooth (a1)-(a3) and striped (b1)-(b3) RWs with $P>0$ and $S>0$. The black dashed lines represent trajectories of positions with the maximum density. For the smooth (striped) rogue waves, $k=-0.5236\approx k_{\mathrm{min}}$ ($k_{1}=-2.963,~k_{2}=2.938$) and $\gamma=1$ ($\gamma=3$). The other parameters are $\epsilon =0.1$, $\eta=1$, $\Omega=1$, $\delta=1$, $\varepsilon=1$, $c_{0}=-1$, and $c_{2}=-1$.}
	\label{fig:analysisrw1k-0}
\end{figure}

For the single NLS equation, the existence of RWs requires the signs of the dispersion coefficient $P$ and nonlinearity coefficient $S$ to be identical, i.e., either $P>0$, $S>0$ or $P<0$, $S<0$. The well-known first-order RW solutions corresponding to these two cases are \cite{AkhmedievN2009Rogue}
\begin{eqnarray}
\varphi _{1}=\frac{\eta }{\sqrt{\pm S}}\left( 1-\frac{4\pm 8\mathrm{i}\eta^{2}T}{1\pm 4\eta^{2}X^{2}/P+4\eta^{4}T^{2}}\right) \mathrm{e}^{\mathrm{i}\left( \omega _{2}\pm \eta^{2}\right) T}.  \label{eq19}
\end{eqnarray}
Substituting these solutions and $X=\epsilon (x-vt)$, $T=\epsilon^2 t$ in equation~\eqref{eq17}, we find that $\omega_{2}$ vanishes. Thus, two types of first-order RWs in the spin-1 BEC with SOC are obtained
\begin{equation}\label{eq20}
\begin{aligned}
\boldsymbol{\psi}=&\frac{\epsilon\eta \mathrm{e}^{\mathrm{i}kx-\mathrm{i}(\omega\mp\epsilon^2\eta^2)t}}{\sqrt{\pm S}} \left(\begin{array}{c}
-\frac{\Omega}{M_1}-\frac{\mathrm{i}(k-v+\gamma)\Omega}{M_1^2}\partial_x \\ 1 \\
-\frac{\Omega}{M_3}-\frac{\mathrm{i}(k-v-\gamma)\Omega}{M_3^2}\partial_x
\end{array}\right)\left( 1-\frac{4\pm 8\mathrm{i}\eta^2\epsilon^2 t}{1\pm 4\eta^{2}\epsilon^2 (x-vt)^2/P+4\eta^{4}\epsilon^4 t^2}\right),
\end{aligned}
\end{equation}
which exhibit positive ($P>0$) and negative ($P<0$) effective masses, respectively. The density profiles $|\psi _{j}|$ of the positive-mass first-order RWs for the case of $P>0$, $S>0$ are shown in figures~\ref{fig:analysisrw1k-0}(a1)-\ref{fig:analysisrw1k-0}(a3).
We find the three components of the spinor state as linearly independent first-order RWs with distinct peak values. Specifically, three first-order RWs exhibit different velocities (the velocities of the position with the maximum density). The reason for this phenomenon is the partial-derivative term in equation~\eqref{eq17}, which is caused by the SOC strength $\gamma$ and the resonant Raman frequency $\Omega $. For the component $\psi_{0}$, the velocity $v_{\psi_{0}}$ is equal to the group velocity $v$, hence $v_{\psi_{0}}=0$ when $k=k_{\mathrm{min}}$. The velocities of $\psi_{+1}$ and $\psi_{-1}$ are, respectively, higher and lower than the velocity of $\psi _{0}$, that is, $v_{\psi _{+1}}<v_{\psi _{0}}<v_{\psi _{-1}}$. For the negative-mass first-order RWs, the velocity relation is reversed. Nonetheless, the maximum peaks of three components always appear at the same spatiotemporal position.

In addition to the RWs $\boldsymbol{\psi }(k)$ shaped like the Peregrine solitons, we can construct striped RWs using a linear combination $\boldsymbol{\psi }(k_{1})+\boldsymbol{\psi }(k_{2})$, where $k_{1}$ and $k_{2}$ satisfy the relation $v(k_{1})=v(k_{2})$. In figures~\ref{fig:analysisrw1k-0}(b1)-\ref{fig:analysisrw1k-0}(b3), we provide an example of a positive-mass striped RW, with $v(k_{1})=v(k_{2})=0$. The striped RWs feature characteristics similar to the smooth ones, except for the density profiles.

\begin{figure}[htbp]
	\centering
	\includegraphics[width=0.7\linewidth]{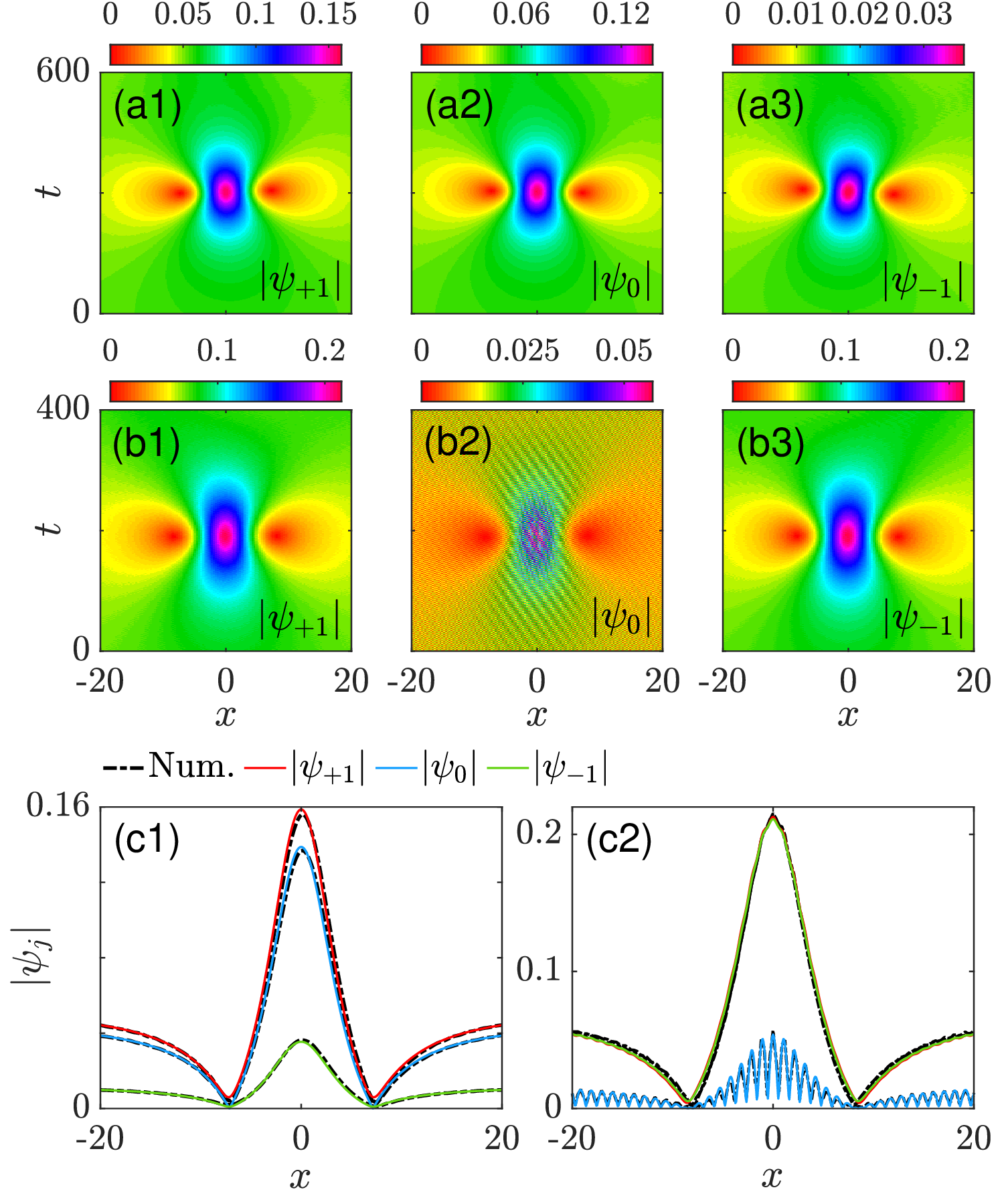}
	\caption{The numerical evolution of the approximate positive-mass first-order smooth (a1)-(a3) and striped (b1)-(b3) RW solutions for $t<0$ with large $|t|$. (c1) and (c2) The comparison of the profiles of the numerical (black dashed curves) and approximate analytical (solid curves) RWs at their maxima. The parameters chosen here are the same as in figure~\ref{fig:analysisrw1k-0}.}
	\label{fig:numericalrw1k-0}
\end{figure}

The RWs in the BEC system exhibit a nonstationary structure with atoms accumulating towards the central portion and then diffusing towards the constant-density background. To check the accuracy of the analytically predicted approximate RW solutions, we simulated the underlying system of the Gross-Pitaevskii equation (\ref{eq3}) using the RW solutions for $t<0$ with large $|t|$ (the moment when RW has not yet appeared). We used the fourth-order Runge-Kutta algorithm in a domain of a large size $[-376\pi <x<+376\pi ]$. to avoid effects of boundary conditions. As shown in figure~\ref{fig:numericalrw1k-0}, we find that the numerically simulated evolution well reproduces the approximate analytical RW solution, with a slight deviation in the velocity of each component. In addition, the wave packet (black dashed curves) at the moment of time corresponding to the highest peak is close to the approximate solution at $t=0$ (solid curves).

\begin{figure}[htbp]
	\centering
	\includegraphics[width=0.7\linewidth]{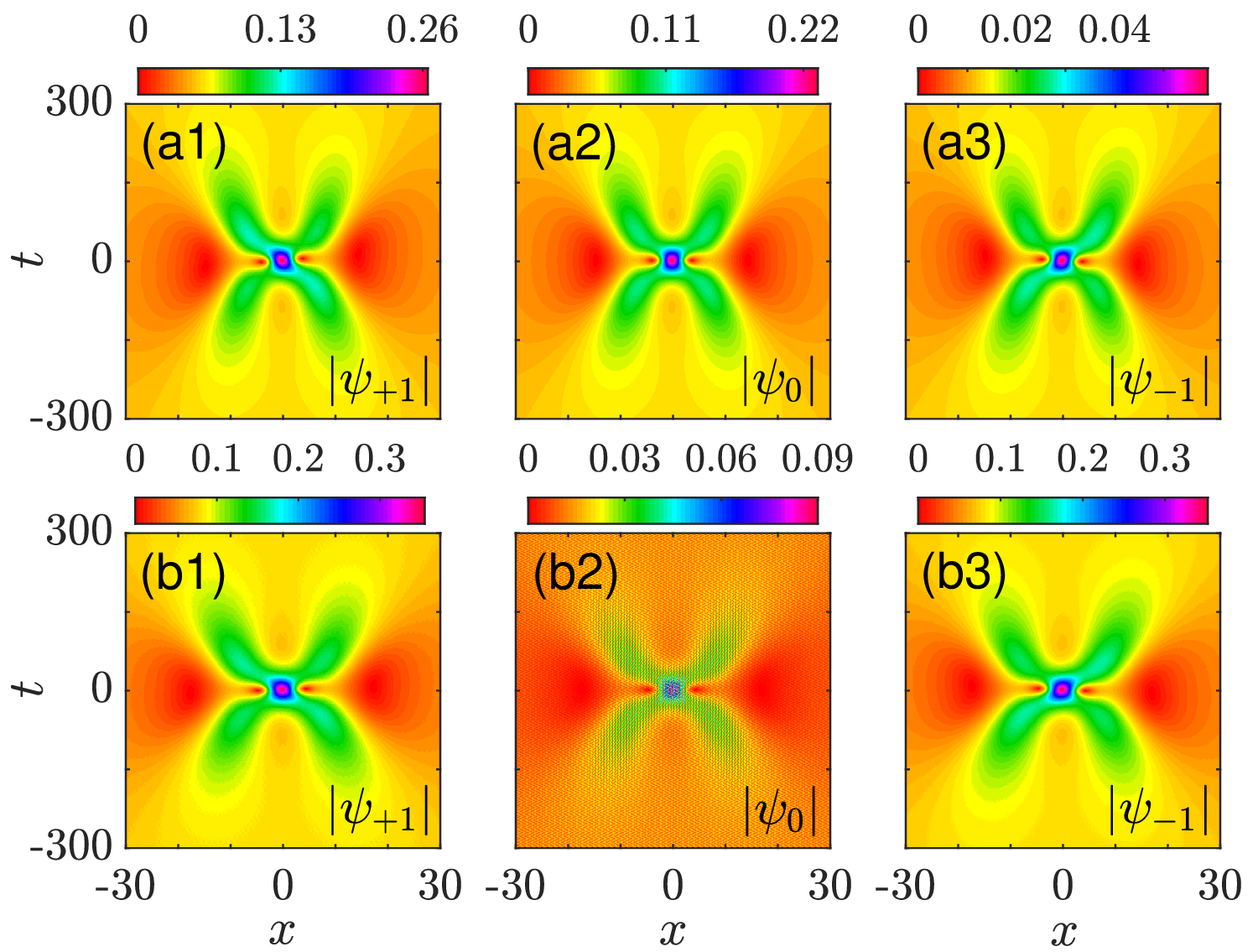}
	\caption{The density profiles of the three-component positive-mass second-order smooth (a1)-(a3) and striped (b1)-(b3) RWs with $P>0$ and $S>0$. The parameters chosen here are the same as in figure~\protect\ref{fig:analysisrw1k-0}.} \label{fig:analysisrw23k-0}
\end{figure}

Similarly, we obtain higher-order RWs in the BEC system (\ref{eq3}) from higher-order RW solutions of the NLS equation~\eqref{eq14}, such as the positive- and negative-mass second-order RW solutions in \ref{appendixA}. The RWs of integer order $r$ are formed by a nonlinear superposition of $r(r+1)/2$ first-order RWs. Taking the positive-mass second-order smooth and striped RWs as examples, shown in figure~\ref{fig:analysisrw23k-0}, we find that each component is shaped as a second-order RW with a primary peak and several secondary ones. Similar to the properties of the first-order vector RWs, the three components of the higher-order RWs are not proportional to each other, and the velocities of the three primary peaks have slight differences. Besides that, the distributions of the secondary peaks are different in the three components. For component $\psi_{0}$, the higher-order RWs have secondary peaks with the same height, like in typical higher-order RWs. On the other hand, the secondary peaks sitting on one diagonal are higher than those on the other one for components $\psi_{\pm 1}$. We have also tested the accuracy of the approximate higher-order RW solutions in numerical simulations. The conclusion is that the numerical solutions corroborate the accuracy of the approximate analytical solutions for the higher-order RWs as well.

\section{Modulation instability and the existence of rogue waves}

\label{sec4} The above analytical solutions indicate that the existence condition for the vector RWs is $PS>0$, in terms of equation~\eqref{eq14}. Therefore, we here focus on the expressions for the effective dispersion and nonlinearity coefficients, $P$ and $S$.

\begin{figure}[htbp]
	\centering
	\includegraphics[width=0.7\linewidth]{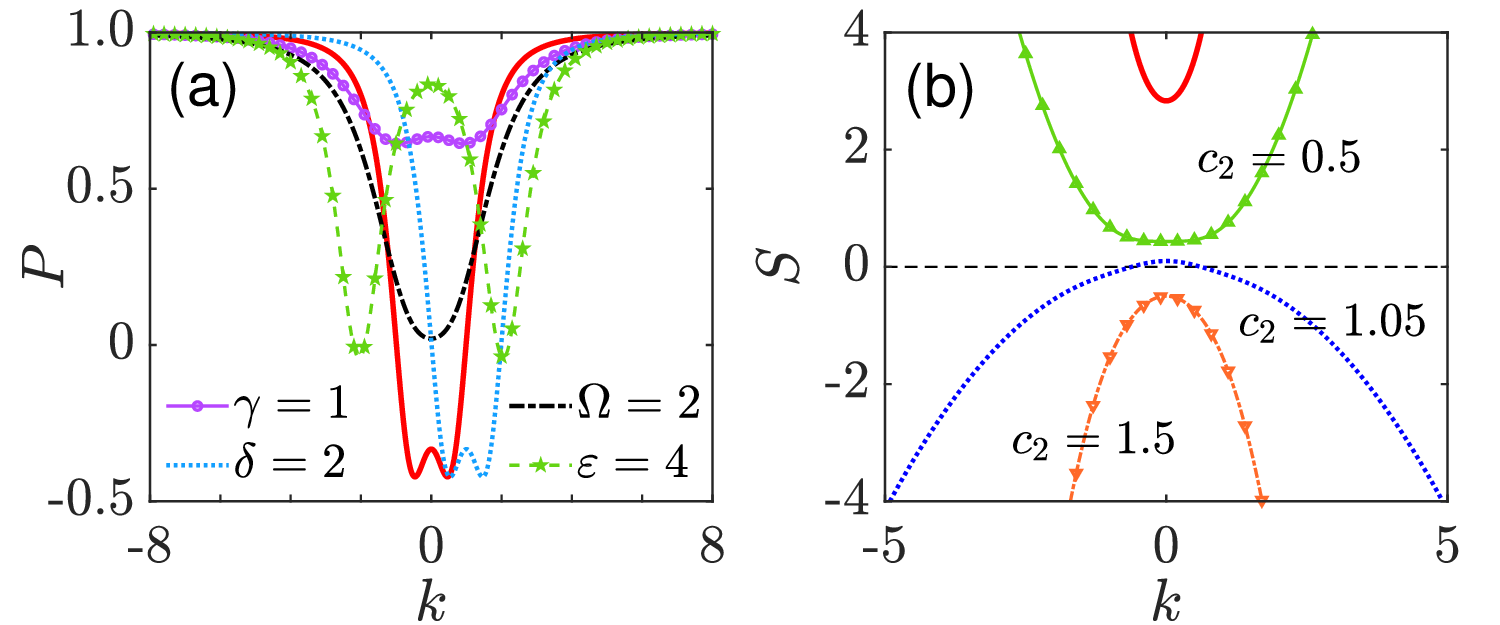}
	\caption{The distribution of (a) the effective dispersion coefficient $P$ and (b) the nonlinearity coefficient $S$ in the momentum $k$-space, for $\gamma=2$, $\Omega=1$, $\delta=0$, $\varepsilon=1 $, $c_{0}=-1$, and $c_{2}=-1$ (red solid lines). Other lines represent the distribution of $P$ or $S$ in the momentum space when only one parameter varies. The horizontal black dashed line in (b) corresponds to $S=0$.}
	\label{fig:pksk}
\end{figure}

A straightforward analysis demonstrates that $M_{1}$ and $M_{3}$ in equation~\eqref{eq9} are positive, thus $U_{1}$ and $W_{1}$ must be negative. Then we derive the effective dispersion $P<1$, which is not affected by nonlinear parameters $c_{0}$ and $c_{2}$, from equation~\eqref{eq15}, and present the results in figure~\ref{fig:pksk}(a). We displayed the distribution of the effective dispersion $P$ in the momentum $k$-space (the red solid line), and the effects of various parameters (the other lines). When $|k|$ is large, the effective dispersion approaches the limit $P\rightarrow 1$. Reducing $\gamma$, increasing $\varepsilon $, or increasing $\Omega $ lead to the increase of the minimum value of $P$ and change the corresponding $|k|$. Detuning $\delta$ does not affect the distribution of $P$ but leads to a shift $\delta /\gamma$ of the distribution in the $k$-space. In addition, we find that the group velocity $v$ with multiple extreme points is a key factor for the occurrence of $P<0$, which requires a relatively large value of $\gamma$.

Next, we shift the focus to equation~\eqref{eq16}. It is obvious that the sign of the nonlinearity coefficient $S$ is bounded by $c_{2}=c_{0}/\beta $, where $-1<\beta=\left( 1-2U_{1}W_{1}\right) ^{2}/\left( 1+U_{1}^{2}+W_{1}^{2}\right) ^{2}-1<0$. In other words, the sign of the coefficient is $S>0$ ($S<0$) when $c_{2}<c_{0}/\beta $ ($c_{2}>c_{0}/\beta $). In the extreme case of $c_{0}<0$ and $c_{2}<-c_{0}$ ($c_{0}>0$ and $c_{2}>-c_{0}$), regardless of the values of $U_{1}$ and $W_{1}$, the sign of $S$ is certainly positive (negative). Figure~\ref{fig:pksk}(b) shows the distribution of $S$ in the momentum $k$-space for different values of $c_{2}$ with $c_{0}<0$. The distribution of $S$ is parabolic and axisymmetric about $k=0$. There is no zero point when $c_{0}$ and $c_{2}$ correspond to the two above-mentioned extreme cases. Conversely, there are two zero points when the two extreme cases do not take place (the blue dotted line in figure~\ref{fig:pksk}(b)), which means that $S$ may be positive or negative.

We obtain the existence domains for the positive- and negative-mass RWs, based on the condition $PS>0$. Because the obtained existence domains are approximate, other methods are required to obtain the existence domains in a more accurate form. A known mechanism for the RW formation is the BMI, which refers to the instability of plane-wave solutions in the baseband limit. In other words, the perturbed evolution of unstable plane waves can generate RWs. Here, we aim to discuss in detail the existence of the vector RWs based on the BMI.

The plane-wave solution of equation~\eqref{eq3} is $\psi_{j}=A_{j}\mathrm{e}^{\mathrm{i}(kx-\tilde{\omega}t)}$, where $A_{j}$ (with $j=\pm 1,0$) and $\tilde{\omega}$ are the amplitude and wavenumber, related by expressions
\begin{equation}
\begin{aligned} \tilde{\omega}=&\frac{k^2}{2} \pm \gamma k \mp \delta+(c_0+c_2)\left(A_{\pm 1}^2+A_{0}^2\right) \\ &+(c_0-c_2)A_{\mp 1}^2+c_2A_0^2\frac{A_{\mp 1}}{A_{\pm 1}}+\Omega \frac{A_0}{A_{\pm 1}},\\ \tilde{\omega}=&\frac{k^2}{2}-\varepsilon+(c_0+c_2)\left(A_{+1}^2+A_{-1}^2\right) \\ &+c_0A_0^2+2c_2A_{+1}A_{-1}+\Omega\left(A_{+1}+A_{-1}\right) .
\end{aligned}
\end{equation}
The complexity of these relations implies that the plane-wave solution of equation~\eqref{eq3} should be studied numerically. For the convenience of the comparison with the above multiscale analysis, we set $A_{0}=1/\sqrt{\pm S}$. Then we add Fourier modes with small amplitudes $a_{j}$ and $b_{j}$ to the plane-wave solution,
\begin{equation}
\psi _{j}=\left[ 1+a_{j}\mathrm{e}^{\mathrm{i}\kappa \left( \lambda t+x\right) }+b_{j}^{\ast }\mathrm{e}^{-\mathrm{i}\kappa \left( \lambda ^{\ast }t+x\right) }\right] A_{j}\mathrm{e}^{\mathrm{i}(kx-\tilde{\omega}t)},  \label{eq21}
\end{equation}
where $\kappa$ denotes a real wavenumber offset from the plane wave, and $\lambda$ is a (complex) eigenfrequency induced by the perturbations. Substituting equation~\eqref{eq21} in equation~\eqref{eq3} and linearizing, we derive a system of six coupled linear equations, $\boldsymbol{L}(a_{+1},b_{+1},a_{0},b_{0},a_{-1},b_{-1})^{T}=0$, where the matrix $\boldsymbol{L}$ is shown in \ref{appendixB}. Obviously, nontrivial perturbation eigenmodes exist under the condition of $\mathrm{det}(\boldsymbol{L})=0$. Eigenvalues of $\lambda$ are obtained from here numerically. If $\lambda$ is a complex number in the baseband limit ($\kappa \rightarrow 0$), BMI occurs, and the corresponding unstable plane wave can excite RWs in the course of its perturbed evolution.

\begin{figure*}[htbp]
	\centering
	\includegraphics[width=0.95\linewidth]{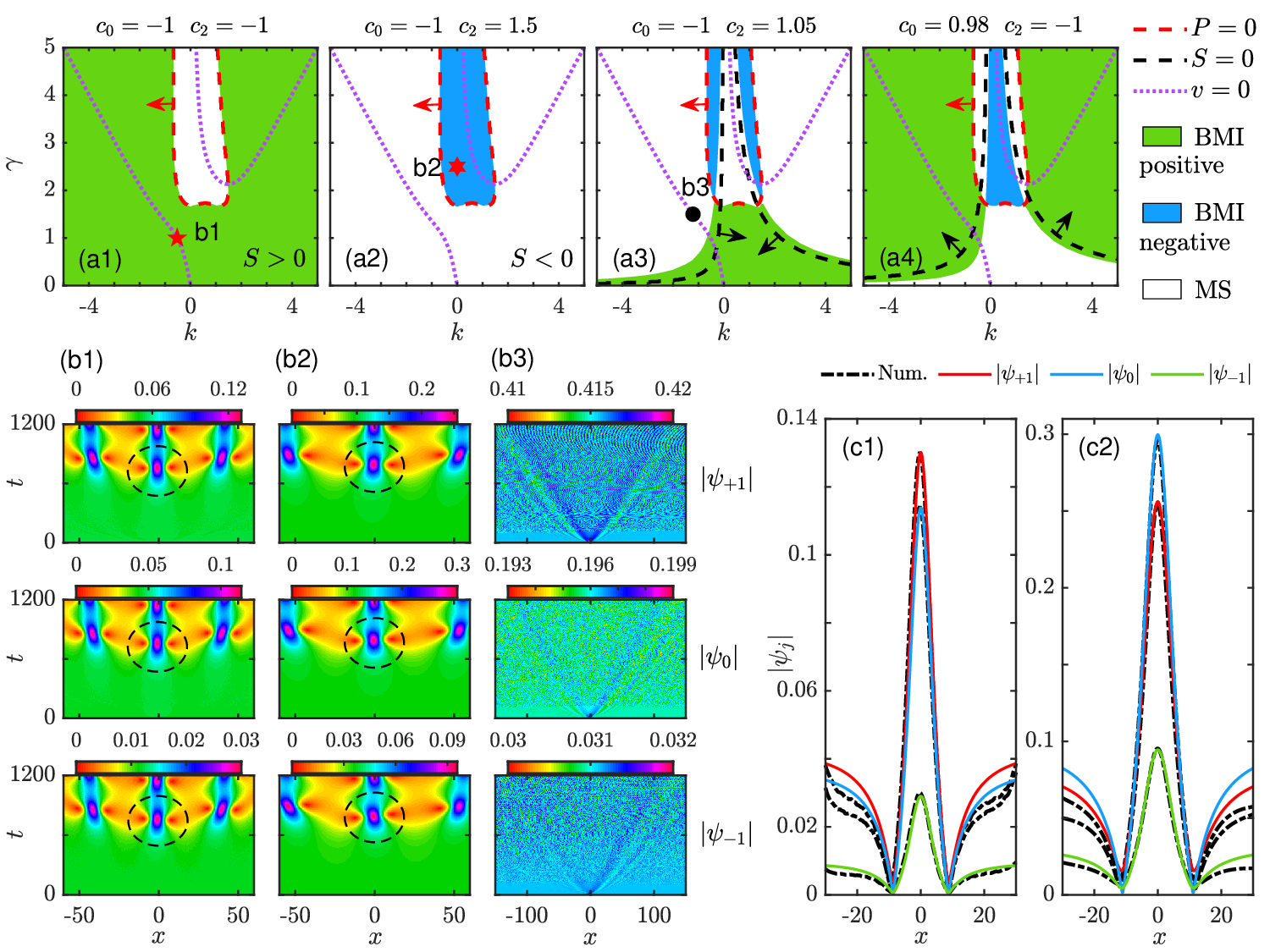}
	\caption{(a1)-(a4) BMI domains in the plane of  ($k$,$\protect\gamma $) for different values of $c_{0}$ and $c_{2}$. Green and blue areas represent the BMI regions for the positive and negative effective masses, respectively. Modulational stability takes place in white areas. The red dashed lines, black dashed lines, and purple dotted ones are contour lines of the vanishing dispersion $P=0$, vanishing nonlinearity $S=0$, and vanishing group velocity $v=0$, respectively. Arrows indicate the direction of gradient increase. The coordinates of the three markers are $(-0.5236,1)$, $(0,2.5)$, and $(-1.216,1.5)$. Here, the background plane wave is $\protect\psi_{j}=A_{j}\mathrm{e}^{\mathrm{i}(kx-\tilde{\protect\omega}t)}$ with $A_{0}=1/\protect\sqrt{\pm S}$, and other parameters are $\protect\epsilon =0.1$, $\Omega =1$, $\protect\delta =1$, and $\protect\varepsilon =1$. (b1)-(b3) The evolution of the plane-wave solutions (corresponding to the three markers) with the $1\%$ Gaussian perturbation. (c1)-(c2) The comparison of the profiles at peaks of the analytical and numerical RWs (circled by black dashed lines), which are excited by the two types of BMI. The background parameters of the analytical RWs for (c1) and (c2) are $\protect\eta=0.82$ and $\protect\eta=0.86$, respectively.}
	\label{fig:mi_and_examples}
\end{figure*}

According to the characteristics of the excited RWs (with the positive or negative effective mass), the BMI can be divided into two types. A generic case can be adequately represented by the BMI of the plane waves in the parameter space ($k,\gamma $) for different $c_{0}$ and $c_{2}$ when $\Omega=1$, $\delta =1$, and $\varepsilon =1$, as shown in figures~\ref{fig:mi_and_examples}(a1)-\ref{fig:mi_and_examples}(a4). In the figures, green (blue) areas are BMI regions for the positive (negative) effective mass. Accordingly, they represent the existence regions for the positive- (negative-) mass RWs. In the same figures. We draw the contour lines corresponding to $P=0$ (red dashed lines) and $S=0$ (black dashed lines), as predicted above by the multiscale method. We find the so predicted RW existence regions, $PS>0$, are very close to the BMI areas determined by the value of $\lambda $, especially the contour lines of $P=0$. The lines of $S=0$ produce a slight deviation, which decreases as the amplitude $\eta $ of the approximate RW solution decreases. Parameters $c_{0}$ and $c_{2}$ affect only the effective interaction coefficient $S$ (determined by the relation between $c_{2}$ and $c_{0}/\beta $), the distribution of the BMI regions changing accordingly. In figures~\ref{fig:mi_and_examples}(a1) and \ref{fig:mi_and_examples}(a2), where the values of $c_{0}$ and $c_{2}$ make the sign of $S$ unique, there is only one type of the BMI (the region with $P>0$ or $P<0$). In figures~\ref{fig:mi_and_examples}(a3) and \ref{fig:mi_and_examples}(a4), both types of the BMI occur because the $S$ may be positive or negative in the plane ($k,\gamma $).

Motivated by the fact that the unstable plane waves can excite RWs in the BMI regions, we numerically verified the BMI regions by means of simulations of the perturbed evolution of the plane-wave solutions $\psi_{j}=A_{j}\mathrm{e}^{\mathrm{i}(kx-\tilde{\omega}t)}$ in the spatial domain $x\in \lbrack -376\pi ,+376\pi ]$ and time interval $0<t<1200$. The numerical results are consistent with our analytical predictions. Figures~\ref{fig:mi_and_examples}(b1)-\ref{fig:mi_and_examples}(b3) show three examples of the simulated evolution of the plane waves with $1\%$ Gaussian perturbations initially added to them, which correspond to the two BMI types one case of modulational stability (MS), respectively. To better observe the structure of the numerically generated RWs, the evolution is displayed in the moving reference frame in which the velocity of component $\psi _{0}$ is set equal to zero. In the BMI regions, we find that the RWs excited in the three components are arranged in an orderly manner. When $P>0$, each RW group with the same space-time coordinates in figure~\ref{fig:mi_and_examples}(b1) exhibits the same shape as the analytical positive-mass RW solutions in figures~\ref{fig:analysisrw1k-0}(a1)-\ref{fig:analysisrw1k-0}(a3), with different peak velocities $v_{\psi_{+1}}<v_{\psi _{0}}<v_{\psi _{-1}}$. For $P<0$, the relation of the velocities is reversed $v_{\psi _{+1}}>v_{\psi _{0}}>v_{\psi _{-1}}$, and the density profiles of the negative-mass RWs are displayed in figure~\ref{fig:mi_and_examples}(b2). In the regions of MS, we find, as expected, that the plane waves with Gaussian perturbations do not generate the RW structure, as shown in figure~\ref{fig:mi_and_examples}(b3). Thus, we verify the existence domains for the positive- and negative-mass smooth RWs. For the striped RWs, an additional condition is that the momenta $k_{1}$ and $k_{2}$ of the linear combination must belong to the BMI region.

We also tried to find approximate analytical solutions corresponding to the numerically found RWs. We have found that the RWs excited by the perturbed plane wave (with the background level $\eta =1$) cannot be described by the analytical RWs with $\eta=1$. Nevertheless, the two BMI-driven evolution examples considered above can be well approximated by the analytical positive- and negative-mass solutions, corresponding to $\eta=0.82$ and $\eta=0.86$, respectively, as shown in figures~\ref{fig:mi_and_examples}(c1) and \ref{fig:mi_and_examples}(c2). This conclusion indicates that the actual background of the excited RW is lower than that of the plane wave which excites the RW, because of the effect of surrounding RWs.

\section{Conclusions}

\label{sec5}We have investigated the vector RW (rogue-wave) solutions in the three-component system for spin-1 BEC with the Raman-induced SOC (spin-orbit coupling). Using the multiscale perturbative method, the underlying system of three-component Gross-Pitaevskii equations was reduced to a single NLS (nonlinear Schr\"{o}dinger) equation. Using well-known RW solutions of the NLS equations, approximate analytical RW solutions for the three-component wave functions were obtained, with the positive ($P>0$) and negative ($P<0$) effective masses. The solutions include RWs with the smooth profile and striped ones, which may be obtained, in the lowest approximation, as a linear combination of smooth ones. The analytically predicted approximate RW states are accurately reproduced by systematic simulations of the underlying system. Solutions for higher-order RWs are also predicted analytically and confirmed numerically. The analytical solutions indicate that an important characteristic of the RWs in the present system is the difference of peak velocities of the RWs in the three components caused by SOC. The velocity relations for the three components of the positive- and negative-mass RWs are reversed. The BMI (baseband modulational instability) of plane waves has been systematically investigated too, revealing RW existence domains. The existence condition $PS>0$ of the analytical RW solutions, where $S$ is the nonlinearity coefficient in the effective NLS equation, provides accurate prediction for the RW existence domains in the underlying spin-1 system. The relevance of the existence domains provided by the BMI analysis was tested by simulations of the perturbed evolution of unstable plane waves. The numerical tests confirm that, in the BMI regions, the evolution of plane waves with small initial Gaussian perturbations readily excites positive- or negative-mass RWs arranged in an orderly manner. The excited RWs are well approximated by the analytical solutions whose background is lower than that of the underlying unstable plane wave. Due to high controllability of all parameters, the vector RWs produced in the present work are likely to be experimentally observed in the spin-1 BEC with SOC, preparing the plane waves in the BMI regions. Our results also indicate that the multiscale perturbative method can be employed to predict various non-stationary solutions of non-integrable models.

\section*{Data availability statement}
All data that support the findings of this study are included within the article (and any supplementary files).

\ack{We acknowledge support of the National Natural Science Foundation of China (Grants No.~11835011, No.~12375006, and No.~12074343) and the NaturalScience Foundation of Zhejiang Province of China (Grant No.~LZ22A050002)}

\appendix
\section{The second-order rogue wave solutions}\label{appendixA}
The positive- and negative-mass second-order rogue wave solutions \cite{AkhmedievN2009Rogue} of the NLS equations \eqref{eq14} are
\begin{equation}\label{appendixeqA1}
\begin{aligned}
\varphi_1=&\frac{\mathrm{e}^{\mathrm{i}\left(\omega_2\pm 1\right)T}}{\sqrt{\pm S}}\left\{1+\left[\frac{3}{8}-\frac{3 X^2}{\pm P}-\frac{2 X^4}{P^2}-T^2\left(9-\frac{12 X^2}{\pm P}\right)-10 T^4
\right.\right.\\ &\left.\left.
\pm \mathrm{i}T\left(\frac{15}{4}+\frac{6 X^2}{\pm P}-\frac{4 X^4}{P^2}-2 T^2-\frac{8 X^2 T^2}{\pm P}-4 T^4\right)\right]/\left[\frac{3}{32}+\frac{9 X^2}{\pm 8P}+\frac{X^4}{2 P^2}
\right.\right.\\ &\left.\left.
+\frac{2 X^6}{\pm 3 P^3}+T^2\left(\frac{33}{8}-\frac{3 X^2}{\pm P}+\frac{2 X^4}{P^2}\right)+T^4\left(\frac{9}{2}+\frac{2 X^2}{\pm P}\right)+\frac{2 T^6}{3}\right]\right\}.
\end{aligned}
\end{equation}
By substituting $X=\epsilon (x-vt)$ and $T=\epsilon^2 t$ into equation \eqref{appendixeqA1} and then substituting the result into equation \eqref{eq17}, the corresponding RWs of the original equation \eqref{eq3} can be obtained.

\section{The matrix for modulation instability analysis}\label{appendixB}
The matrix $\boldsymbol{L}$ related to modulation instability is expressed as
\begin{equation}
\boldsymbol{L}=\left(\begin{array}{cccccc}
-L_{1}-\Lambda_1 & -c_+A_{+1}^2 & L_{13} & -c_+A_{0}^2 & -c_-A_{-1}^2 & L_{16} \\
c_+A_{+1}^2 & L_{1}-\Lambda_1 & c_+A_{0}^2 & L_{24} & L_{25} & c_-A_{-1}^2 \\
L_{31} & -c_+A_{+1}^2 & -L_2-\Lambda_2 & L_{34} & L_{35} & -c_+A_{-1}^2 \\
c_+A_{+1}^2 & L_{42} & L_{43} & L_{2}-\Lambda_2 & c_+A_{-1}^2 & L_{46} \\
-c_-A_{+1}^2 & L_{52} & L_{53} & -c_+A_{0}^2 & -L_{3}-\Lambda_3 & -c_+A_{-1}^2 \\
L_{61} & c_-A_{+1}^2 & c_+A_{0}^2 & L_{64} & c_+A_{-1}^2 & L_{3}-\Lambda_3
\end{array}\right),
\end{equation}
where the diagonal elements are
\begin{equation}
\begin{aligned}
L_{1}&=\left( k^{2}+\kappa^{2}\right) /2+\gamma k-\delta -\omega+c_{+}(2A_{+1}^{2}+A_{0}^{2})+c_{-}A_{-1}^{2}, \\
L_{2}&=\left( k^{2}+\kappa^{2}\right) /2-\varepsilon -\omega+c_{+}(A_{+1}^{2}+A_{-1}^{2})+2c_{0}A_{0}^{2}, \\
L_{3}&=\left( k^{2}+\kappa^{2}\right) /2-\gamma k+\delta -\omega+c_{+}(2A_{-1}^{2}+A_{0}^{2})+c_{-}A_{+1}^{2}, \\
\Lambda_{1}&=(k+\gamma+\lambda )\kappa, ~~\Lambda_{2}=(k+\lambda )\kappa,~~\Lambda_{3}=(k-\gamma +\lambda )\kappa,
\end{aligned}
\end{equation}
and off-diagonal ones are
\begin{equation}
\begin{aligned}
L_{24}&=-L_{13}=c_{+}A_{0}^{2}+2c_{2}A_{0}^{2}A_{-1}/A_{+1}+\Omega A_{0}/A_{+1},\\
L_{64}&=-L_{53}=c_{+}A_{0}^{2}+2c_{2}A_{0}^{2}A_{+1}/A_{-1}+\Omega A_{0}/A_{-1},\\
L_{43}&=-L_{34}=c_{0}A_{0}^{2}+2c_{2}A_{+1}A_{-1},\\
L_{31}&=-L_{42}=c_{+}A_{+1}^{2}+2c_{2}A_{+1}A_{-1}+\Omega A_{+1}/A_{0},\\
L_{61}&=-L_{52}=c_{-}A_{+1}^{2}+c_{2}A_{0}^{2}A_{+1}/A_{-1},\\
L_{25}&=-L_{16}=c_{-}A_{-1}^{2}+c_{2}A_{0}^{2}A_{-1}/A_{+1},\\
L_{46}&=-L_{35}=c_{+}A_{-1}^{2}+2c_{2}A_{+1}A_{-1}+\Omega A_{-1}/A_{0}.
\end{aligned}
\end{equation}
Here, $c_{\pm }=c_{0}\pm c_{2}$.

\end{document}